\pdfoutput=1
\documentclass[prl,twocolumn,tightenlines,showpacs,amsfonts,amssymb,amsmath]{revtex4-1}
\usepackage{amsmath}
\usepackage{bm}
\usepackage{natbib}
\usepackage{url}
\usepackage{graphicx}
\usepackage[breaklinks,colorlinks,
  urlcolor=blue,citecolor=blue,linkcolor=blue]{hyperref}
\begin{document}

\newcommand{\dd}{\mathrm{d}}
\newcommand{\ee}{\mathrm{e}}
\newcommand{\ii}{\mathrm{i}}
\newcommand{\der}{\partial}
\newcommand{\vev}[1]{\langle#1\rangle}

\def\ga{\mathrel{\raise.3ex\hbox{$>$\kern-.75em\lower1ex\hbox{$\sim$}}}}
\def\la{\mathrel{\raise.3ex\hbox{$<$\kern-.75em\lower1ex\hbox{$\sim$}}}}
\def\bk{{\mathbf k}}
\def\bp{{\mathbf p}}
\def\bn{{\mathbf n}}
\def\bx{{\mathbf x}}


\title{Anisotropies of Gravitational Wave Backgrounds: A Line Of Sight
  Approach.}

\author{Carlo R. Contaldi}
\affiliation{
Theoretical Physics, Blackett Laboratory, Imperial College, London, SW7 2AZ, UK  
}

\begin{abstract}
  In the weak field regime, gravitational waves can be considered as
  being made up of collisionless, relativistic tensor modes that
  travel along null geodesics of the perturbed background metric. We
  work in this geometric optics picture to calculate the anisotropies
  in gravitational wave backgrounds resulting from astrophysical and
  cosmological sources. Our formalism yields expressions for the
  angular power spectrum of the anisotropies. We show how the
  anisotropies are sourced by intrinsic, Doppler, Sachs-Wolfe, and
  Integrated Sachs-Wolfe terms in analogy with Cosmic Microwave
  Background photons.
\end{abstract}
 \maketitle

{\sl Introduction.} The first direct detection of gravitational waves
\cite{PhysRevLett.116.241103} has heralded a new era of gravitational
wave astronomy. Future improvements in ground \cite{Aasi:2013wya} and
space-based detectors \cite{PhysRevLett.116.231101,LISA} and pulsar
timing arrays promise to increase sensitivity to the level required to
carry out detailed studies of stochastic and relic backgrounds of
gravitational waves. Pulsar timings arrays, in particular, probe the
longest baselines providing a window at frequencies not accessible to
ground and space-based detector (see
eg. \cite{Shannon:2015ect,2010CQGra..27h4013H,2015aska.confE..37J}). Advanced,
space-based interferometer detectors such as the proposed Big Bang
Observer \cite{Crowder:2005nr,PhysRevD.73.083511} will target the same
frequency window as ground based detectors but may reach the
sensitivity required to detect the relic background left over from
inflation.

Stochastic backgrounds of gravitational waves are made up of the
superposition of astrophysical signals from unresolved sources. A
number of different source mechanisms may result in stochastic
backgrounds including the merger of compact objects, emission from cosmic
string networks or phase transitions in the early universe (see
eg. \cite{Binetruy:2012ze,AmaroSeoane:2012km}. Although
these sources can be at cosmological distances we differentiate them
from the relic, or cosmological, background due to inflation or another mechanism
operating at much higher redshifts. The different backgrounds can be
distinguished, in principle, by their different frequency scalings and
statistical properties \cite{Regimbau:2011rp}.

The energy flux of any gravitational wave background will not be
constant across the sky. These anisotropies will contain information
about the mechanism that generated the gravitational waves and about
the nature of the spacetime along the line of propagation of the
waves. If backgrounds will be detected in future it is interesting to
consider what anisotropic signal we should expect. Various proposals
have been made for how to map anisotropies in the
backgrounds \cite{Cornish:2001hg,Mitra:2007mc,Thrane:2009fp,Romano:2015uma}. However,
little progress has been made to calculate the expected
anisotropy. This is distinct from the calculation of the anisotropy
due to clustering of resolved sources \cite{Namikawa:2015prh} which
may be of interest in luminosity distance constraints. An isotropic
background has angular distance dependence but no luminosity distance
dependence.

Although extremely challenging measurements of even just the largest
angular scales of the anisotropies will provide interesting
astrophysical and/or cosmological constraints. If anisotropies in the
relic background were measured directly they would provide a unique
window onto the Planckian epoch. 

In this {\sl letter} we introduce a new formalism for calculating
gravitational wave background anisotropies. Our assumption will be
that some future detectors will have the sensitivity to determine the
energy flux in the gravitational wave background as a function of
direction in the sky. We start by developing a Boltzmann equation for
the perturbation in the distribution function of the tensor, or
``graviton'', modes propagating the energy flux of the gravitational
waves. This is in analogy with Cosmic Microwave Background (CMB)
calculations. We extend this analogy further by employing the
line-of-sight method to obtain a form for the angular power spectrum
of the anisotropies that consist of a time integral of generic
Legendre expanded source functions. We show how the formalism can be
employed to calculate the signal from different generic emission
mechanisms responsible for the backgrounds.

Our approach takes into account both source and line-of-sight
effects and shows how gravitational waves could,
in principle, be used as a non-electromagnetic probe of the universe over
cosmological distances. This is very counter-intuitive since it
suggests using tensor perturbations to probe scalar-dominated
perturbations to the background metric. However the astonishing
sensitivity being forecast in the field of direct detection of
gravitational waves make this a concrete, albeit long-term, possibility.

{\sl Graviton distribution function.} Gravitons, like photons, are
assumed to be massless modes, or fluctuations, that make up the energy and momentum
flux carried by the coherent oscillations known as gravitational or
electromagnetic waves respectively. In analogy with photons we can
treat of gravitons as propagating along null geodesics of the
background spacetimes. We must be careful in making this assumption
since gravitons, unlike photons, are a direct manifestation of the
perturbation of the spacetime in a non-linear theory. However, the
shortwave formalism, developed over fifty years
ago \cite{Misner:1974qy}, shows that this is a good approximation in
the weak field limit {\sl even} when the curvature of the background
is large. In this picture we can take a geometric optics approach to
the propagation of gravitational waves by considering them as made up
of a stream of massless, collisionless, gravitons following null
geodesics. The null geodesics are determined by the perturbed background.

The energy flux carried by gravitational waves is conserved. In the
geometric optics approach this is included by defining an adiabatic
invariant such as the graviton number density, or phase-space
distribution function. This is the starting point for our calculation,
in direct analogy with that of CMB anisotropies~\footnote{We
follow closely the notation and pedagogical itinerary
of \cite{Dodelson:2003ft} throughout.}.

At zeroth order in perturbations, the energy and the magnitude of its
three momentum of a massless graviton are set by a single parameter
$p$ that is also proportional to the frequency $\nu$. For
gravitational waves emitted isotropically and homogeneously, with an energy
spectrum $dE/d\nu$, the distribution function will
be a function of time $t$ and frequency only
\begin{equation}\label{eq:distfunc}
f(\nu,t) = \frac{1}{\nu^3}\frac{dE}{d\nu}\,.
\end{equation}

The total energy density carried by a gravitational wave can be
obtained by integrating the distribution function over the three
momentum using the definition of the infinitesimal momentum volume
element $p^2dpd\,\Omega \equiv \nu^2d\nu\,d\Omega$, where
$d\Omega$ is the momentum space infinitesimal angular element. For the
isotropic case we then have
\begin{equation}
\rho_{\rm gw}(t) \equiv \int d\nu\, d\Omega\, \nu^3\,f(\nu,t) =4\pi\int d\nu \frac{dE}{d\nu}\,.
\end{equation}

The spectrum $dE/d\nu$ is specific to the mechanism generating the
gravitational waves (see eg. \cite{Regimbau:2011rp} for a review) and
$\rho_{\rm gw}$ is related to the {\sl strain power} measured by
detectors, related to the square of the wave amplitude. Alternatively
$f(\nu,t)$ can be considered as proportional to the the {\sl specific
intensity} of the gravitational waves \cite{Mitra:2007mc}. The
polarization of the gravitational waves is of great interest but for
simplicity we will assume the measurement is {\sl not} polarization
sensitive in this work. 

{\sl Anisotropies.} We now introduce anisotropies by allowing the
distribution function to depend on the arrival direction $\hat \bn$ with $f\equiv
f(\nu,{\hat \bn},t)\equiv
f(p,{\hat \bn},t)$. The anisotropies are due to inhomogeneities in
either the source mechanism or the propagation of the gravitaional
waves. We consider
first order perturbations around a Friedmann-Robertson-Walker (FRW)
metric with scale factor $a(t)$ and coordinates~\footnote{We use units where $c=h=G=1$ in the
  following.} $x^\mu =
(c\,t,\bx)$. The metric in the Newtonian gauge, with scalar, first
order perturbations, is given by; $g_{00} =
-(1+2\Psi)$, $g_{ij} = \delta_{ij}
a^2(t)(1+2\Phi)$, and $g_{0i}= g_{i0} = 0$, 
with $\Psi(\bx)$ and $\Phi(\bx)$ the first order scalar potential and curvature
perturbations.

The four momentum of the gravitons is defined with respect to the affine
parameter $\lambda$ along the particle's trajectory $P^\mu = dx^\mu/d\lambda$.
The energy of the massless graviton is now perturbed and we have
$P^2= -(1+2\Psi)(P^0)^2 + p^2 = 0$, 
with three momentum magnitude $p^2 = g_{ij}P^iP^j$. Then at first
order in the perturbation we have $P^0 = p\,(1-\Psi)$ and
$P^i = p\,\hat p^i(1-\Phi)/a$,
where components $p^i$ define the instantaneous unit vector for the propagation.

Liouville's theorem states that $df/d\lambda=0$ in the absence of
collisions and injection of modes.  Adding collision and source
operators, ${\cal C}[...]$ and ${\cal J}[...]$, on the right hand side
of Liouville's equation, we obtain a Boltzmann--type equation 
\begin{equation}\label{eq:liouville}
  \frac{df}{d\lambda} = {\cal C}[f(\lambda)] + {\cal J}[f(\lambda)]\,.
\end{equation}

In the case of gravitational waves the collisional term
is not present. The emission term is present however and for astrophysical
source it will be {\sl extended} in time. This is distinct from the 
CMB case where collisions are present until last scattering and
the injection is included simply as thermal initial conditions.

The left hand side of (\ref{eq:liouville}), given the perturbed metric,
 can be expanded using the perturbed geodesic
equation and re-written in terms of physical time $t$ \cite{Dodelson:2003ft}.

For the right hand side of (\ref{eq:liouville}) we introduce a source
term of the form $df/dt = j(t) f$ 
defined by an emissivity rate per comoving volume $j(t)$.

The distribution function itself must be expanded in the
perturbations. On the left hand side of the Boltzmann equation we
consider perturbations in the energy of the modes via an expansion to
first order in a dimensionless perturbation~\footnote{In CMB
  calculations this would be the $\Delta T/T$ around the reference
  Planckian spectrum of thermal photons.} $\Gamma(\bx,\hat{\bp},t)$
\begin{equation}\label{eq:flhs}
  f(p[1+\Gamma]) \approx f(p) + p\frac{\partial f}{\partial
    p}\Gamma\,.
\end{equation}

To expand on the the emission side we introduce a perturbation due to the
peculiar velocity $v(\bx,t)$ of the emitter, with respect to
the rest frame of the observer, and an inhomogeneity in the emission $\Pi(\bx,t)$
\begin{equation}\label{eq:frhs}
   f(p[1+\hat p^iv_i+\Pi]) \approx f(p) + p\frac{\partial f}{\partial
    p}\left(\hat p^iv_i+\Pi\right) \,.
\end{equation}
Notice that we chosen to introduce the inhomogeneity of emission as an
inhomogeneous perturbation of the energy per mode rather than an
inhomogeneity in the emissivity. Our choice simplifies the formalism
and the two give equivalent perturbations to the energy density of the
gravitational waves. We have also assumed that the emission mechanism is still
isotropic apart from the Doppler--like term due to the peculiar
velocity of the emitter.

Inserting these into~(\ref{eq:liouville}) we obtain a zeroth order equation
\begin{equation}\label{eq:back}
  \frac{\partial f}{\partial t} - H \, p\frac{\partial
    f}{\partial p} =  j\,f \,,
\end{equation}
where $H$ is the background Hubble rate. This describes the redshifting
of the spectrum of the gravitaional waves due to the expanding
background and the growth of the monopole of the background due to any
time dependent emission mechanism. In essence,
integrating~(\ref{eq:back}) determines $\rho_{\rm gw}(\nu)$.

At first order, after rearranging and expanding in plane waves with
wavevectors $\bk$ with
$\bk\cdot\hat\bp=k\,\mu$ and changing to conformal time $\eta$, we
obtain a differential equation for the dimensionless perturbation
\begin{equation}\label{eq:anis}
  \dot\Gamma + \left(i\,k\mu+\dot\sigma\right)\Gamma =
  \dot\sigma\left( \hat
  p^iv_i+\Pi \right) + \dot\Phi +i\,k\mu\,\Psi\,,
\end{equation}
where an over dot represents a derivative with respect to $\eta$ and
we have introduced the {\sl conformal emissivity rate}
$\dot\sigma \equiv a\,j$. The perturbation does not depend on $p$ so
the anisotropies will have the same frequency dependence as the
monopole $\rho_{\rm gw}(\nu)$.

{\sl Streaming of gravitational waves.} Equation~(\ref{eq:anis})
describes the evolution of the anisotropy in the specific intensity of
gravitational waves given their streaming along perturbed geodesics
{\it and} the injection of waves with a given spectrum and rate. Its
form is intentionally similar to the equivalent equation for CMB anisotropies.

We now use the line-of-sight integration
method \cite{Seljak:1996is} to determine the anisotropy at our
location today,  $\eta=\eta_0$ by integrating (\ref{eq:anis}). Just
as with CMB calculations, we can make use of the fact that the
directional dependence is determined purely by the inner product of
the gravitational wave momentum vector $\bp$ (the line-of-sight) with the plane
wavevector $\bk=k\,{\hat \bk}$.
The $\mu$ dependence can be isolated through
integration by parts and the perturbation can be Legendre expanded to
obtain a multipole expansion of the anisotropy
\begin{equation}\label{eq:gammal}
  \Gamma_\ell(k,\eta_0) = \int^{\eta_0}_{\eta_i} \,d\eta\,j_\ell\left[k(\eta_0-\eta)\right]\,e^{-\Delta\sigma}\,S(k,\eta)\,.
\end{equation}
Here $\eta_i$ is an initial time, $\Delta\sigma(\eta) \equiv
\sigma(\eta_0)-\sigma(\eta)$, $j_\ell$ are spherical Bessel functions,
and we have assumed that
$\Gamma(\eta_i)\to 0$. The direction
{\sl independent} source function is given by
\begin{equation}\label{eq:source}
S(k,\eta) = \dot\Phi-\dot\Psi+\dot\sigma\left({\hat p}^iv_i+\Pi-\Psi\right)\,.
\end{equation}

This expression is the main result of this work. Each term
in~(\ref{eq:source}) is due to well understood physical effects with
counterparts in the CMB source function.  The
first two terms and last term in the brackets are the Integrated Sachs-Wolfe
(ISW) and Sachs-Wolfe (SW) effects
respectively \cite{1967ApJ...147...73S,1968Natur.217..511R,Sugiyama:1994ed}.
The remaining terms are a Doppler contribution due to the peculiar velocity
of the emitter and an intrinsic contribution due to the inhomogeneous
distribution of emitters. The emissivity rate $\dot \sigma$ defines an
emission ``depth'' in analogy to the optical depth parameter $\tau$
for CMB photons.

The SW effect arises from the gravitational redshift caused by the local
curvature at emission. This effect is somewhat ambiguous for the case
of gravitational waves since there may be strong, non-linear effects
from the dynamics involved
in the emission mechanism but we may interpret it as the effect of the
local curvature perturbation in the asymptotic spacetime at a certain
distance from the source. In the following we will assume vanishing
anisotropic stresses in the scalar perturbations by setting $\Psi=-\Phi$.

A gravitational wave transfer function can be defined by dividing the
perturbation by the primordial, scalar curvature perturbation
$\Delta^h_\ell(k,\eta_0) = \Gamma_\ell(k,\eta_0)/\Phi_0(k)$. It may
seems unnatural to normalize the modes by the {\sl scalar} amplitude
but we have done this in anticipation that only in the case of a relic
background would the primordial tensor amplitude appear in the
emission contribution to the source function. The appearance of a
primordial tensor amplitude can always be accounted for by using
the primordial tensor-to-scalar ratio $r$.



By considering the spherical harmonic coefficients of the perturbation
$a_{\ell m}^h(\eta_0)$ we can obtain an expression for the angular
power spectrum of the gravitational wave background anisotropies
\begin{equation}\label{eq:spectrum}
C^{h}_\ell=\frac{2}{\pi} \int k^2 dk P_\Phi(k)|\Delta_\ell^{h}(k,\eta_0)|^2\,,
\end{equation}
where we have introduced the power spectrum of primordial curvature
perturbations $k^3P_\Phi(k) = A_s k^{n_s-1}$.


\begin{figure}[t]
\centerline{
\includegraphics[width=0.5\textwidth]{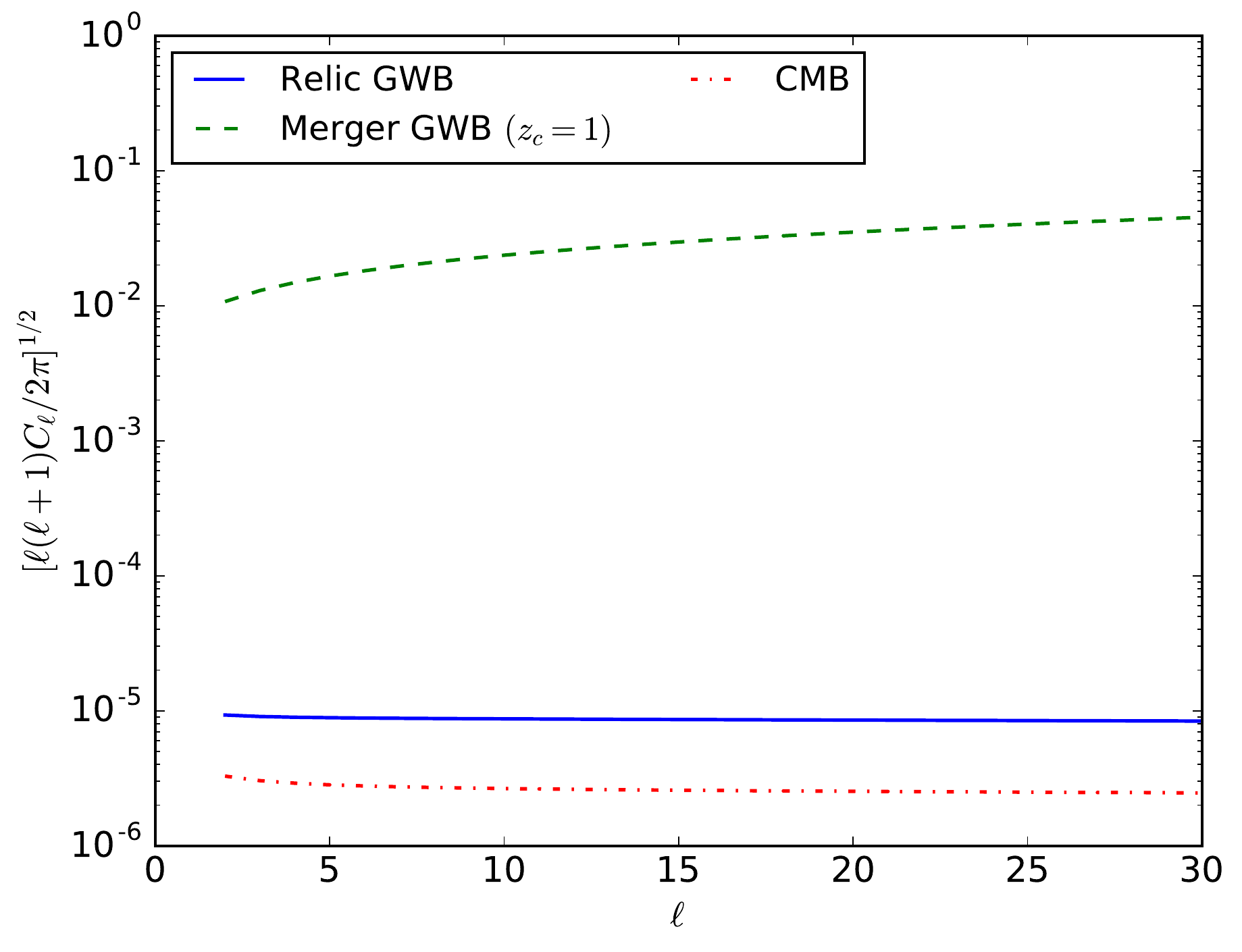}
} \caption{Angular power spectrum of anisotropies in $\rho_{\rm gw}$
for merger and relic backgrounds and the CMB for standard
inflationary primordial spectra. The merger model uses a simple
merger rate peaking at redshift $z=1$ to model the emission.
} \label{fig:clh}
\end{figure}

{\sl Anisotropies from compact object mergers.} For backgrounds
arising from mergers of compact objects such as black hole collisions
(BHBH) or black hole, neutron star (BHNS) collisions it is reasonable
to assume that the perturbation to the density of sources is a biased
tracer of the perturbation to the background matter with the form
$\Pi(\vec k,\eta) = b(k,\eta)\delta_m(\vec k,\eta)$ where $b\sim{\cal
O}(1)$ is the linear bias parameter and $\delta_m$ is the dark matter density
contrast. The statistics of the dark matter
distribution is determined by the late-time matter power spectrum
$\vev{\delta^{\,}_m(\vec k,\eta_0)\delta^\star_m(\vec
    k',\eta_0)}=(2\pi)^3 \delta(\vec k - \vec k')P(k,\eta_0)$.

The matter power spectrum can itself be related to
the spectrum of primordial curvature perturbations via the Poisson
equation $k^2\Phi(\vec k,\eta) = 4\pi\, a^2\delta_m\rho_m$. Thus when
considering correlations in the anisotropy $\Gamma$ we should be able to
relate these to the statistics of the underlying curvature
perturbation, albeit via a heavily biased tracer in the case of merger
sources. The Doppler term in~(\ref{eq:source}) can also be related to
the primordial curvature via linear perturbation theory but we shall
omit it here for simplicity. 

We can now use linear growth of structure to relate the primordial and
late-time curvature perturbations using $\Phi(\vec k,\eta)
= \frac{9}{10}\Phi_0(\vec k)T(k)g(\eta)$. Here $T(k)$ is the matter
transfer function \cite{Bardeen:1985tr} with $T\to 1$ on large scales
and $g(\eta)$ is the scale independent growth function.
Inserting these relations into~(\ref{eq:source}) we obtain the source function
\begin{equation}\label{eq:merger}
S(k,\eta) =
\frac{3}{5}T(k)g(\eta)
\left[\left(\frac{b\, k^2a}{\Omega_m
    H_0^2}+\frac{3}{2}\right)\dot\sigma+3\frac{\dot g}{g} \right]\,,
\end{equation}
with $\Omega_m$ the matter energy density in units of the critical
energy density. The rate $\dot\sigma$ is determined by the merger rate
history of each class of sources.

The dominant term in~(\ref{eq:merger}) is the intrinsic one. This is
determined by the fact that emission peaks at relatively low redshifts
for any merger mechanism. At these redshifts even the smallest $k$
probed by the project on the largest angular scales have ratios
$k^2a/\Omega_mH_0^2\gg 1$. This is not surprising; even if
we remove resolved sources we are still left with a background
dominated by shot noise statistics driven by clustered sources.

{\sl Relic background anisotropies.} As long as we are in the weak
field regime, the formalism can be extended to super horizon scales
and applied to relic gravitational waves produced during a period of
inflation. In this case the intrinsic term in the source function is
just the sum of the primordial amplitudes of the two independent gravitational
wave polarizations $h_+$ and $h_\times$. The emissivity rate
$\dot\sigma$ in this case becomes a delta function at
$\eta_i$ which is assumed to be some time where all the modes of
interest are far outside the horizon. The source function then takes
the form
\begin{equation}
S(k,\eta)= \frac{9}{5}T(k)\dot g(\eta)+\left(h_+ + h_\times + \Phi_0\right)\,,
\end{equation}
where we have, once again, dropped the Doppler term for simplicity.
When employing this in~(\ref{eq:spectrum}) we can make use of the
power law form for the spectrum of initial tensor perturbations giving
$k^3P_{h_{+,\times}}(k) = r\, A_s \, k^{n_t}/2$.

The transfer function for relic backgrounds takes on the particularly
simple form on large scales with $T(k)=1$
\begin{equation}
\Delta_\ell^h(k,\eta_0) \!=\!
(1+r\,k^{n_t})j_\ell\left[k(\eta_0-\eta_i)\right]+\frac{9}{5}\!\!\int^{\eta_0}_{\eta_i}\!\!\!
\dot g\,
j_\ell [k(\eta_0-\eta)]d\eta\,.
\end{equation}
In this case contributions from the combined SW and intrinsic and
ISW effects will be of comparable magnitude on large angular scales,
if the ISW is present, as with the CMB case.

It is instructive to compare with the CMB 
photon transfer function on large angular scales, assuming
instantaneous recombination at $\eta=\eta_\star$
\begin{equation}
\Delta_\ell^\gamma(k,\eta_0) \!=\!-\frac{3}{10}\,j_\ell[k(\eta_0-\eta_\star)]-\frac{9}{5} \!\!\int^{\eta_0}_0
\!\!\!\dot g\,
j_\ell [k(\eta_0-\eta)]e^{-\tau}d\eta\,.
\end{equation}
For an inflationary background the CMB and gravitational wave
background anisotropies will be highly correlated. The relic
anisotropies will have a larger amplitude than the CMB since
gravitational waves ``last scattered'' long before radiation to matter
transition. The CMB also suffers from attenuation from
rescattering. The anisotropies from low redshift mergers are orders of
magnitude larger than either relic or CMB anisotropies but we should
still expect a correlation on large scales from ISW and SW terms
although this will be difficult to model precisely in practice.

In Fig.~\ref{fig:clh} we show the $C_\ell$ obtained from the
expressions above for a standard $\Lambda$CDM model with power-law
primordial spectra. For the merger background we use a very simple toy
model for the merger rate that peaks at redshift $z=1$ and use a bias
factor $b=3$. The relic background uses $r=0.1$ and $n_t=0.16$.

{\sl Discussion.} We have introduced a line-of-sight approach for the
calculation of anisotropies in gravitational wave backgrounds. The
feasibility of future measurements of this kind remains unclear but it
is important to quantify the information in these signals.  Our
result shows how anisotropies are induced by inhomogeneities at the
source and along the line-of-sight. We have used our expressions to
make some preliminary estimates of the anisotropies about the monopole
in the backgrounds but further work is needed to
model the signal accurately, particularly for the compact source case. 

{\sl Acknowledgements.} We thank Jo\~ao Magueijo and particularly Toby
Wiseman for very useful discussions. This work was supported by an
STFC grant number ST/L00044X/1.

\bibliography{refs}

\end{document}